\documentclass[12pt,preprint]{aastex}

\shorttitle{Nature of Dark Energy}
\shortauthors{Solevi et al.}

\begin{document}

\title{The Nature of Dark Energy from deep Cluster Abundance}     

\author{P.~Solevi, R.~Mainini \& S.A.~Bonometto}
\affil{Physics department `G.~Occhialini', University of
Milano--Bicocca, Piazza della Scienza 3, I20126 Milano, Italy \&
INFN, Sezione di Milano, Via Celoria 16, I20133 Milano, Italy}
\begin{abstract} 
We show that using the redshift dependence of the
deep cluster abundance to detect the nature of Dark Energy
is a serious challenge. We compare the expected differences
between flat $\Lambda$CDM models, with different $\Omega_{mo}$, 
with the difference between $\Lambda$CDM and dynamical DE
models. In the former case, cluster
abundances in comoving volume and geometrical factors
act in the same direction, yielding a significant difference
between the expected angular densities. On the contrary,
when we keep a constant $\Omega_{mo}$ and change the DE nature,
abundances in comoving volume and geometrical factors
act in the opposite direction, so that the expected differences
in angular densities reduce to small factors.
\end{abstract}

     
\section{Introduction}

High redshift supernovae, data on the cosmic microwave background
(CMB), as well as on the large-scale galactic distribution (Riess et
al. 1988, Perlmutter et al 1988, Tegmark et al. 2001, De Bernardis et
al 2000, Hanany et al 2000, Halverson et al 2001, Spergel et al 2003,
Percival et al. 2002, Efstathiou et al 2002) indicate that $\sim 70\,
\%$ of the world contents are due to a smooth component with negative
pressure, such that the ratio $w \equiv p/\rho < \sim -0.8$. The
nature of this compenent, dubbed dark energy (DE), is still open
for debate. Candidates range from false vacuum, yielding a positive
cosmological constant $\Lambda$, to a self--interacting scalar field
$\phi$ (Ratra \& Peebles 1988; Wetterich 1988) to even more exotic
physics of extra dimensions (e.g., Dvali \& Turner 2003).

While the observed value of $\Lambda$ implies a dramatic {\it
fine--tuning} of vacuum, at the latest phase transition, $\Lambda$CDM
models provide an excellent fit to data. Alternative viable DE models
must, first of all, rival their success. Therefore, their
phaenomenology must be and {\it is} hardly distinguishable from
$\Lambda$CDM.  This calls for tests able to discriminate between
different DE natures.

In recent work, the evolution of the cluster mass function has been
shown to have a significant dependence on DE nature. This has been
shown first on the basis of a Press \& Schechter (PS) formulation (Mainini,
Macci\`o \& Bonometto 2003), then using n--body simulations (Klypin et
al. 2003; see also Linder \& Jenkins 2003), which confirmed PS
findings.

There are a number of other cosmological measures which can contribute
to discriminate between different DE natures. This is important also
on the light of the point that we wish to make in this paper. In fact,
while the number density of clusters $n(>M,z)$, in a comoving volume,
has a significant $z$ dependence, the measurable signal is much less
significant. On the light of observational uncertainties, it seems
unlikely that the nature of DE can be easily discriminated along this
pattern.

In a sense, this contradicts the expectations for a number
of deep sky surveys (see, e.g., Davis, Gerke \& Newman 2004)
based on the ancient intuition of Hubble (1926), that the
number--redshift relation can be used to determine the geormetry of
the world. 
Discriminating the DE nature is much harder than fixing the matter
density parameter $\Omega_m$ in a flat $\Lambda$CDM model.

It is not so because of a reduced impact of DE nature on geometry or
on $n(>M,z)$, but because their effects on geometry and $n(>M,z)$ tend
to erase each other.

\section{Geometrical effects}

Let us consider a family of objects whose (cumulative) mass function
is $n(>M,z)$. Indipendently of the actual $z$ dependence, the angular
number density of objects belonging to such family, with redshift
between $z$ and $z+ \Delta z$, in a spatially flat geometry, reads
\begin{equation}
N(>M,z,\Delta z) = \int_{z}^{z+\Delta z} dz'\, D(z')\, r^2(z')\, n(>M,z')
\label{eq:1}
\end{equation}
with $D(z) = dr/dz$. For the models we are considering, it is useful
to show that
\begin{equation}
D(z) = {c \over H_o} \sqrt{ \Omega_m(z) \over \Omega_{mo} (1+z)^3 }
\label{eq:2}
\end{equation}
and, therefore,
\begin{equation}
r(z) = {c \over H_o} \int_o^z dz' 
\sqrt{ \Omega_m(z') \over \Omega_{mo} (1+z')^3 }~.
\label{eq:3}
\end{equation}
In fact, for flat models, on the past light cone, $a\, dr = -c\, dt$
and, therefore, by dividing the two sides by $dz=-da/a^2$, we obtain
$a\, dr/dz = a^2 c\, dt/da$, so that:
\begin{equation}
D(z) = {dr \over dz} = {c \over H(z)}~~~~~~~~ \bigg({\rm here,~as~usual,~}H = 
{\dot a \over a}\bigg).
\label{eq:4}
\end{equation}
In turn
$$
\, ~H^2 (z) = {8\pi \over 3} G {\rho_m (z) \over \Omega_m(z)}
= {8\pi \over 3} G {\rho_{mo} (1+z)^3 \over \Omega_m(z)} = ~
$$
\begin{equation}
~~~~~~~~~~~~~~~~~
= H_o^2 { \Omega_{mo} (1+z)^3 \over \Omega_m(z)}~,
\end{equation}
so that eqs.~(\ref{eq:2}) and (\ref{eq:3}) follow.

These equations lead to handable expressions when DE has a
state equation $w = p/\rho$ with constant $w$.
It is then easy to see that
\begin{equation}
D(z) = (c/ H_o) (1+z)^{-{3/ 2}} [\Omega_{mo} +
(1-\Omega_{mo})(1+z)^{3w}]^{-1/ 2},
\label{eq:6}
\end{equation}
while $r(z)$ can be worked out by integrating from 0 to $z$.
We can also easily differentiate both with respect of $\Omega_{mo}$ and
with respect to $w$, finding that
\begin{equation}
{\partial D^2 \over \partial \Omega_{mo}} = -\big({c \over  H_o}\big)^2
{1-(1+z)^{3w} \over (1+z)^3 
[\Omega_{mo} + (1-\Omega_{mo})(1+z)^{3w}]^{2}}
\label{eq:7}
\end{equation}
and
\begin{equation}
{\partial D^2 \over \partial w}
=
-\big({c \over  H_o} \big)^2
{(1-\Omega_{mo})3\ln(1+z)\, (1+z)^{3w} \over (1+z)^3 
[\Omega_{mo} + (1-\Omega_{mo})(1+z)^{3w}]^{2}}
~,
\label{eq:8}
\end{equation}
so that we expect that the geometrical factors increase when 
$\Omega_{DEo} = 1-\Omega_{mo}$ increases, whenever $w<0$,
and decrease when $w$ increases. Notice that eq.~(\ref{eq:8}) is 
not a functional derivative, as the expression (\ref{eq:6})
holds just for constant $w$.

\begin{figure}      
\plotone{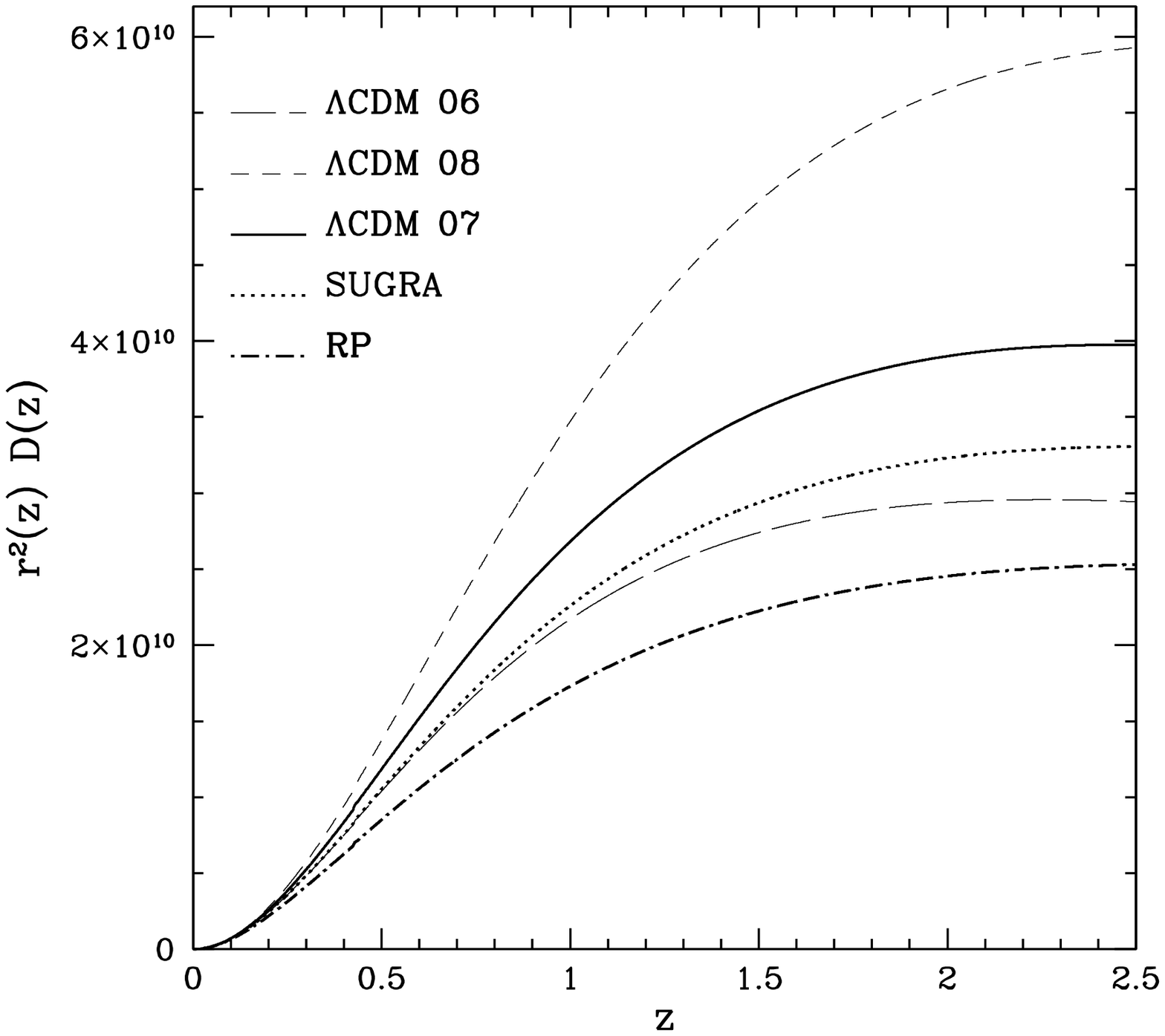}	
\epsscale{1.0}
\caption{Redshift dependence of geometrical factors 
for $\Lambda$CDM and dynamical DE models.} 
\label{fig:geo}
\end{figure}

An extension to dynamical DE can be however performed by using the
interpolating expressions yielding $\Omega_m(z)$, for RP (Ratra \&
Peebles, 1988, 1995) and SUGRA (Brax \& Martin 1999, 2001, Brax,
Martin \& Riazuelo 2000) models, provided by Mainini et al (2004).  In
these cases DE is due to a scalar field, self--interacting through a
potential whose expression depends on an energy scale $\Lambda$ (see
Appendix A for details on the dynamical DE models considered here).
Using the interpolating expressions or, equivalently, direct numerical
integration, we obtain the results shown in Figure \ref{fig:geo}. Here
the $z$ dependence of the geometrical factor $D(z) r^2(z)$ is shown
for three $\Lambda$CDM models ($\Lambda$CDM 06, $\Lambda$CDM 07,
$\Lambda$CDM 08, with $\Omega_{mo} = 0.4$, 0.3 and 0.2,respectively),
as is obtainable from eq.~(\ref{eq:6}). In the same Figure we show the
$z$ dependence of geometrical factors also for SUGRA and RP models,
with $\Omega_{mo} = 0.3$ and with the $\Lambda$ parameter fixed at
10$^3$GeV.  $H_o$ is 70$\, $km/s/Mpc in all models.

In the absence of number density evolution, Fig.~1 would also
show the dependence of the angular number density on $z$.

The $\Lambda$CDM models considered are characterized by quite
different values of $\Omega_{mo}$. Among them, only 0.7 approaches
available data, which are coherent with an $\Omega_{mo}$ interval not
wider than 0.02--0.03.  In spite of that, the discrepancy between $\Lambda$CDM
07 and $\Lambda$CDM 06 only marginally exceeds the discrepancy between SUGRA
and $\Lambda$CDM 07 and is smaller than the difference between RP and 
$\Lambda$CDM 07.
Notice, in particular, that the RP model can be considered as
discrepant from data as $\Lambda$CDM 06 or $\Lambda$CDM 08.

Altogether, Fig.~1 shows how strongly the nature of DE affects
geometrical factors and that, in the presence of a non--evolving
population, an insight into the DE equation of state can be provided
by the $z$ dependence of their angular number density.

\section{Cluster number evolution}

A fair insight into the evolution of the number of clusters with the
redshift $z$, can be obtained by using a PS expression. Sheth \&
Tormen (1999, 2002) as well as Jenkins et al (2001) provided
expressions more closely fitting n--body simulations (which are
already reasonably approached by PS results). The latter expressions
are more complex and include more parameters, while their use is
unessential for the present aims.

The expected (differential) cluster number density $n(M)$,
at a given time, is then given by the expression
\begin{equation}
f(\nu) \nu d\log \nu
= {M \over \rho_m} n(M) M d\log M~.
\label{eq:2-1}
\end{equation}
Here $\rho_m$ is the matter density, $\nu = \delta_c/\sigma_M$ is the
so--called {\it bias factor}, $M$ is the mass scale considered.
$\sigma_M$ is the r.m.s. density fluctuation on the scale $M$ and
$\delta_c$ is the amplitude that, in the linear theory, fluctuations
should have in order that, assuming spherical evolution, full
recollapse is attained exactly at the time considered (in a
standard CDM model this value is $\sim$1.68; the difference, in
other model, ranges around a few percent).
As usual, we took a Gaussian $f(\nu)$ distribution.

Together with eq.~(\ref{eq:2-1}), we must take into account
the virialization condition, which yields significantly
different density contrasts $\Delta_c$ in different DE models.
Further details can be found in Mainini et al (2003).

\begin{figure}      
\plotone{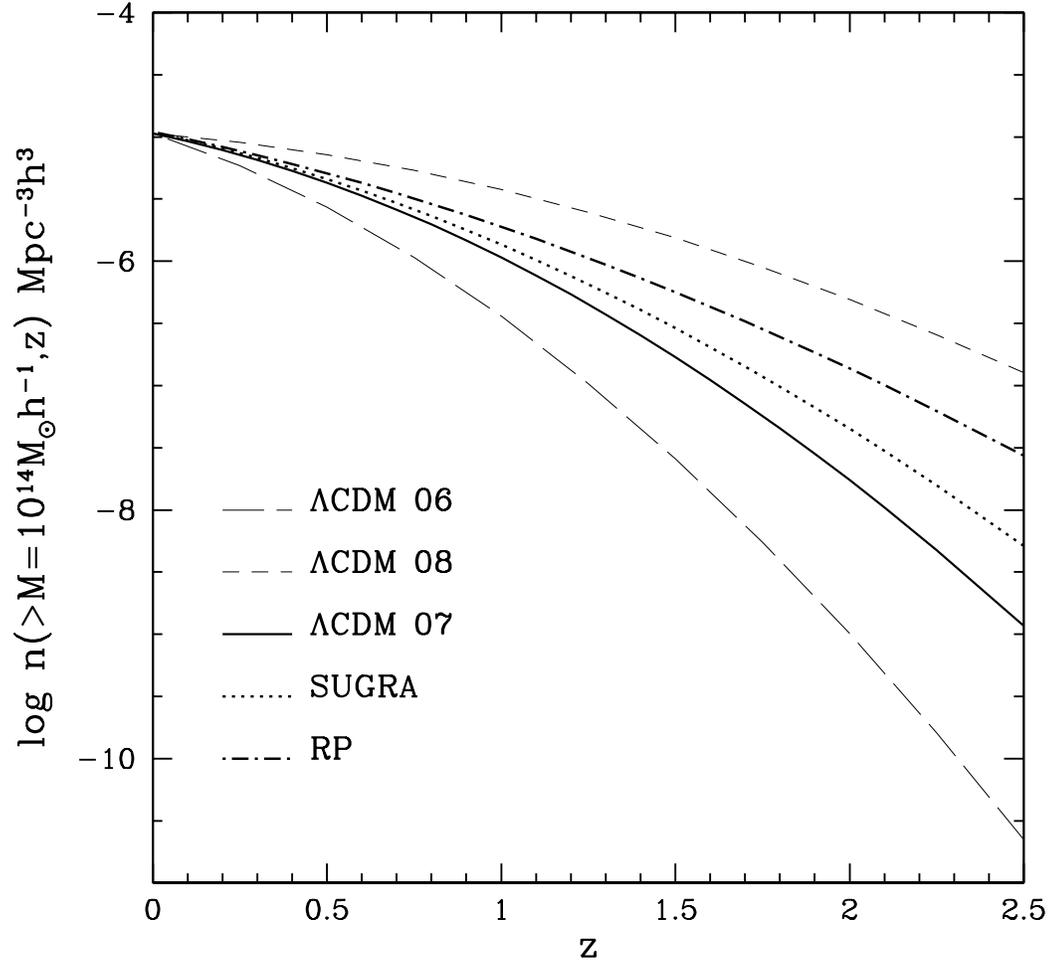}	
\epsscale{1.0}
\caption{Cluster number, in comoving volumes, in $\Lambda$CDM and 
dynamical DE models. In this and in the following plots $\Delta z = 0.1$.} 
\label{fig:numden}
\end{figure}

In Figure \ref{fig:numden} we show the cumulative cluster number
density, $n(>M,z)$, obtained by integrating $n(M)$, for the same
models of Fig.~1$\, $, for $M = 10^{14} M_\odot h^{-1}$. All models
are normalized to the same cluster number today and the redshift
dependence of $n(>M,z)$ is clearly understandable, on qualitative
bases: When $\Lambda$CDM models are considered, the evolution is
faster as we approach standard CDM. Hence, $\Lambda$CDM 08
($\Lambda$CDM 06), being the most distant (nearest) model to standard
CDM, has the slowest (fastest) evolution. $\Lambda$CDM 07 just stands
in between.

When we compare dynamical DE models to $\Lambda$CDM 07, we know that
they also yield a slower evolution, {i.e.} structures form
earlier in models with $w > -1$. 

Accordingly, apart of numerical details, the behaviors shown
in Fig.~2 reasonably fit our expectations.

\section{Conclusions}

Let us now put together the results shown by Figs.~1~\&~2 and evaluate
the angular number densities of clusters in redshift intervals
$z,~z+\Delta z $ with $\Delta z = 0.1$.  If $\Lambda$CDM models are
considered, the geometrical factor $D(z) \, r^2(z)$ and the
evolutionary factor $n(>M,z)$ vary in the same directions, when
$\Omega_{mo}$ is modified.

On the contrary, for dynamical DE models, the two terms vary in the
opposite directions and, therefore, model differences tend to erase.

\begin{figure}      
\plotone{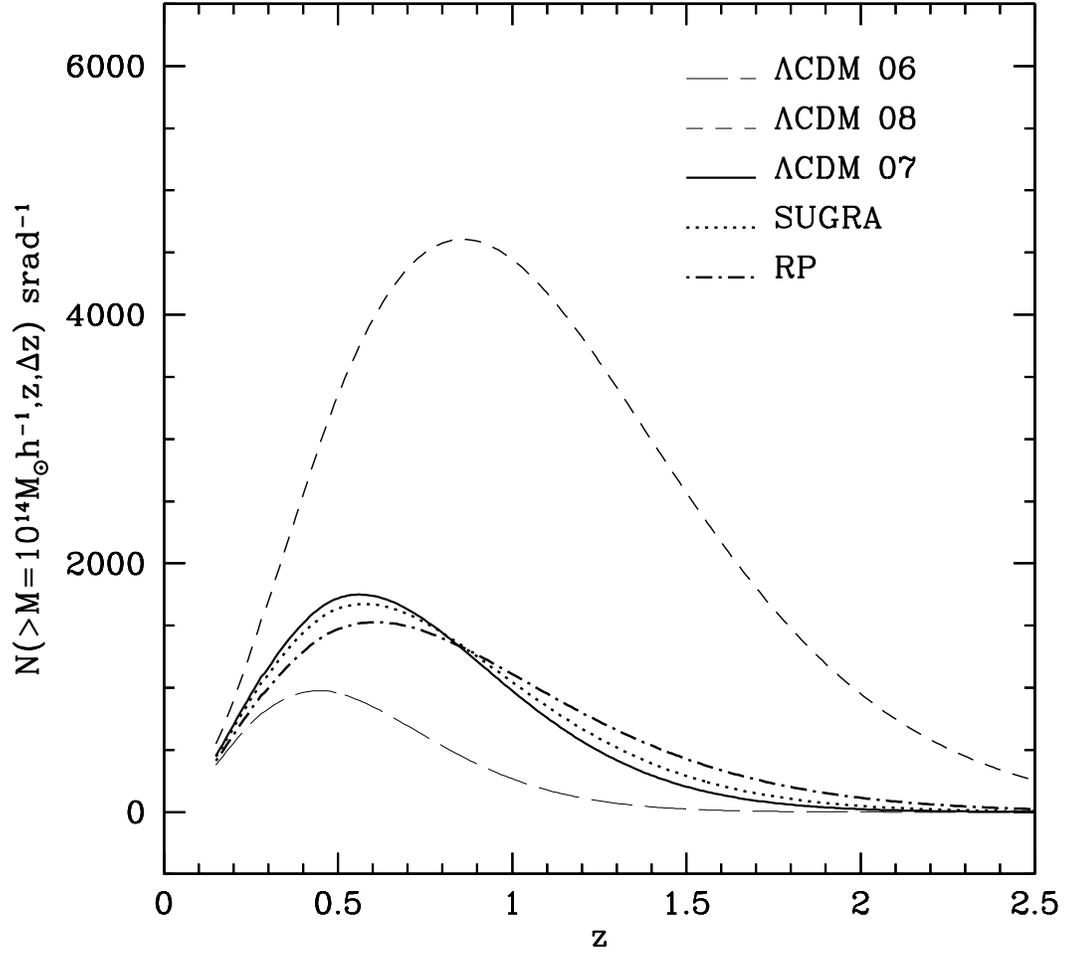}	
\epsscale{1.0}
\caption{Cluster number, on the celestial sphere, in $\Lambda$CDM and
dynamical DE models.} 
\label{fig:area}
\end{figure}

\begin{figure}      
\plotone{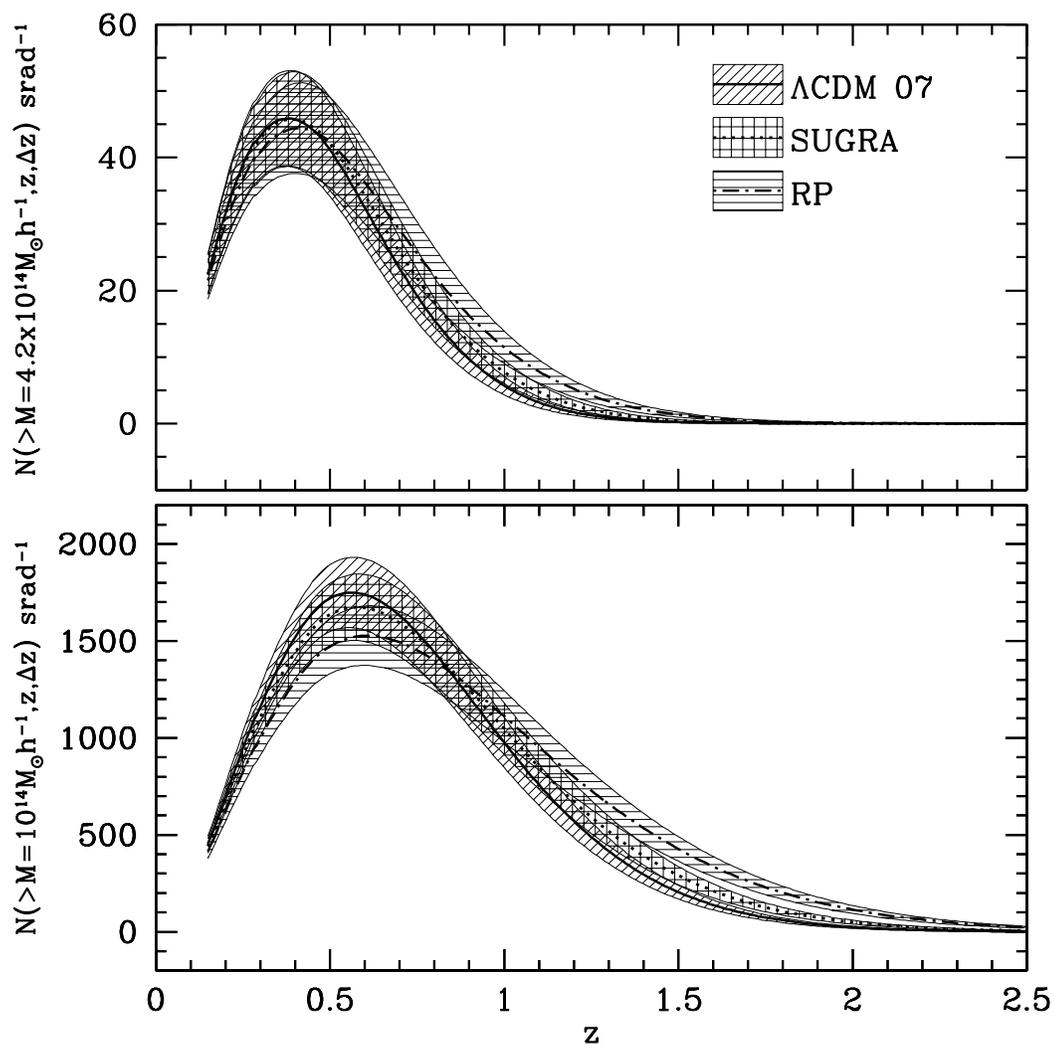}
\epsscale{1.0}	
\caption{Cluster number, on the celestial sphere, in $\Lambda$CDM and
dynamical DE models with $\Omega_{mo}=0.7$. Effects due to an uncertainty
of $5 \%$ in the mass determination are also shown. In the range
$1<z<2$ and for $M=10^{14}h^{-1} M_\odot$, discrimination may be
possible after suitable improvements of data and theory.} 
\label{fig:inc}
\end{figure}

Accordingly, the results shown in Figure~\ref{fig:area} are
essentially expected. It is then easy to argue that testing the
differences between cluster numbers is really challenging, when
different DE models are compared. On the contrary, different
$\Omega_{mo}$'s cause comparatively huge shifts.  Data on cluster
masses are obtainable either from temperatures $T$ or luminosities $L$
(see, e.g., Pierpaoli et al 2004).
Finding a fair cluster mass is perhaps the main reason for uncertainty, 
but also the cluster redshift determination is subject to errors.
In Figure~\ref{fig:inc} we compare the expected mass functions, for
the $\Lambda$CDM 07 and the two dynamical DE models considered, at two
mass scales ($10^{14} M_\odot h^{-1}$ and $4.2 \cdot 10^{14} M_\odot
h^{-1}$), showing also the uncertainty  in the halo mass function 
caused by a 5$\, \%$ indetermination in the mass:
$\delta N = 
[\, M n(M)/ N(> M, z, \Delta z)\,]~ (\delta M/ M)~,
$
still for $\Delta z=0.1$.
Of course, $\delta M/ M \sim 0.05$ is not easily reachable, on the
basis of the present capacity to reconstruct cluster dynamics. The
plots also assume that a complete cluster sample in a solid angle of 1
steradian is available.

Using deep cluster distributions, to discriminate DE models, is
therefore hard. A comparison between the two panels of
Figure~\ref{fig:inc} however shows that differences become wider as we
go to smaller masses. Accordingly, on group mass scales, a further
enhancement can be expected. An even stronger difference can be
expected, when angular number densities of galaxies are being
compared. In particular, comparing the two panels shows that, when
lowering the mass scale considered, differences in trends
(vs.~redshift) are enhanced. On galaxy mass scales, therefore, the
very peaks of distributions could fall at differend redshifts.
Predictions for galaxy scales cannot be formulated only using a PS
recipe, but should include an accurate consideration of how large
halos split into galactic objects. This will be the topic of future
research.

Before concluding let us however outline that, here above, the dependence
on $z$ of cluster features has been assumed not to depend on DE nature.
As a matter of fact, when we consider different state equations for
DE, we change the $z$--$t$ relation and the same redshift corresponds
to a different time. This is true also when $\Omega_{mo}$ is
changed, of course. However, while, in the latter case, this change
of the $z$--$t$ relation has a counterpart in a change of halo
angular densities, there is no equivalent counterpart in the
former case. Comparing the $z$--$t$ relation, obtained from a
study of cluster morphologies or galactic evolution, with 
number density predictions can be therefore a way to discriminate
between different DE natures.

\acknowledgments
Andrea Macci\'o and Loris Colombo are  gratefully thanked for
interesting comments. 
\appendix
\section{Dynamical DE models}

Dynamical DE is to be ascribed to a scalar field, $\phi$,
self--interacting through an effective potential $V(\phi)$, whose
dynamics is set by the Lagrangian density:
\begin{equation}
\label{eq:lag}
{\cal L}_{DE}  = -{1\over2} \, \sqrt{-g} \, 
\left( \partial^\mu \phi \partial_\mu \phi + V(\phi) \right)~.
\end{equation}
Here $g$ is the determinant of the metric tensor $g_{\mu  \nu}= 
a^2(\tau) dx_{\mu} dx_{\nu}$ ($\tau $ is the conformal
time). In this work we need to consider just a spatially homogeneous
$\phi$ ($\partial_i \phi \ll
\dot \phi$; $i = 1,2,3$; dots denote differentiation with respect
to $\tau$); the equation of motion then reads:
\begin{equation}
\ddot \phi +2{\dot a \over a} \dot \phi+ a^2 {dV \over {d \phi}} = 0~.
\label{eq:motion} 
\end{equation}
Energy density and pressure, obtained from the
energy--momentum tensor $T_{\mu \nu}$, are:
\begin{equation}
\label{eq:prho}
\rho = -T^0_0 =  {\dot \phi^2\over {2a}} + V(\phi)~,
~~~~~
p ={1\over 3} \, T^i_i = {\dot \phi^2\over {2a}} - V(\phi)~,
\end{equation}
so that the state parameter
\begin{equation}
\label{eq:wde}
w \equiv {p \over \rho} ={
{ {\dot\phi^2/{2a}} - V(\phi)}\over { {\dot\phi^2/{2a}} 
+ V(\phi)}}
\end{equation}
changes with time and is negative as soon as the potential term $V(\phi)$ takes
large enough values.

The evolution of dynamical DE depends on details of the effective
potential $V(\phi)$. Here we referred to
models proposed by Ratra \&
Peebles (RP: 1988, 1995), yielding
a rather slow evolution of $w$, and 
Brax \& Martin (SUGRA: 1999, 2001, see also Brax,
Martin \& Riazuelo 2000) yielding a much faster evolving $w$.
Altogether, RP and SUGRA potentials cover a large spectrum of evolving
$w$. They read
\begin{equation}
V(\phi) = \frac{\Lambda^{4+\alpha}} {\phi^\alpha} \qquad RP,
\label{eq:1a}	
\end{equation}
\begin{equation}
V(\phi) = \frac{\Lambda^{4+\alpha}}{\phi^\alpha} \exp (4\pi G
\phi^2)~~~ SUGRA.
\label{eq:2a}	
\end{equation}
These potentials allow tracker solutions, yielding the same low--$z$
behavior, almost independently of initial conditions.
In eqs.~(\ref{eq:1a}) and (\ref{eq:2a}) , 
$\Lambda$ is an energy scale, currently set in the range
$10^2$--$10^{10}\, $GeV, relevant for the physics of fundamental
interactions. The potentials depend also on the exponent $\alpha$.
Fixing $\Lambda$ and $\alpha$, the DE density parameter $\Omega_{DE}$
is determined. Here we rather use $\Lambda$ and $\Omega_{DE}$ as
independent parameters. In particular, numerical results are given
for $\Lambda=10^3$GeV.

The RP model with such $\Lambda$ value is in slight disagreement with low-$l$
multipoles of the CMB anisotropy spectrum data. Agreement may be recovered
with smaller $\Lambda$'s, which however loose significance in particle
physics. The SUGRA model considered here, on the contrary, is in fair
agreement with all available data.


\end{document}